\documentclass[12pt, oneside]{article}   	
\usepackage{jheppub}
\usepackage{amsmath}
\usepackage{amssymb}
\usepackage{amsfonts}
\usepackage{amsthm}
\usepackage{amsbsy}
\usepackage{array}
\usepackage{mathtools}
\usepackage[usenames,dvipsnames]{xcolor}
\usepackage{subcaption}
\usepackage{graphicx}
\usepackage[vcentermath]{youngtab}

\makeatletter
\def\@fpheader{\ }
\makeatother

\newcommand\snowmass{\begin{center}\rule[-0.2in]{\hsize}{0.01in}\\\rule{\hsize}{0.01in}\\
\vskip 0.1in Submitted to the  Proceedings of the US Community Study\\ 
on the Future of Particle Physics (Snowmass 2021)\\ 
\rule{\hsize}{0.01in}\\\rule[+0.2in]{\hsize}{0.01in} \end{center}}

\title{Snowmass Topical Summary: Formal QFT}
\author[a]{David Poland,}
\author[b]{Leonardo Rastelli}
\affiliation[a]{Department of Physics, Yale University, New Haven, CT 06520, USA}
\affiliation[b]{C. N. Yang Institute for Theoretical Physics, Stony Brook University, Stony Brook, NY 11794-3840, USA}

\emailAdd{david.poland@yale.edu}
\emailAdd{leonardo.rastelli@stonybrook.edu}

\abstract{We attempt to give a broad conceptual overview of modern quantum field theory, highlighting important recent developments. This report serves as the TF03 topical group summary for Snowmass 2021.}

\notoc
\begin{document}

\maketitle
\vspace{2cm}
\noindent We are grateful for the community input we have received, especially from the authors of white papers submitted to TF03:

\vspace{.25cm} 
\noindent {\it Philip C. Argyres, Ibrahima Bah, Clay Cordova, Mykola Dedushenko, Thomas T. Dumitrescu, A. Liam Fitzpatrick, Daniel Freed, Rajesh Gopakumar, Sarah M. Harrison, Thomas Hartman, Jeffrey A. Harvey, Jonathan J. Heckman, Kenneth Intriligator, Emanuel Katz, Martin Kruczenski, Mario Martone, Dalimil Maz\'{a}\v{c}, Gregory W. Moore, Nikita Nekrasov, Natalie M. Paquette, Sabrina Pasterski, Monica Pate, Jo\~ao Penedones, Eric Perlmutter, Silviu S. Pufu, Ana-Maria Raclariu,  Shlomo S. Razamat, Sakura Sch\"{a}fer-Nameki, Shu-Heng Shao, David Simmons-Duffin, Balt C. van Rees, Xi Yin, Alexander Zhiboedov}

\vspace{2cm}
\snowmass

\pagenumbering{roman}
\setcounter{page}{2}
\newpage
\pagenumbering{arabic}
\setcounter{page}{1}

\section*{Executive summary}

Quantum field theory (QFT) sits at the heart of modern physics, serving as a universal framework for both particle and condensed matter physics. Despite its immense utility, the question of what QFT {\it is} -- how to properly conceptualize it and define it --  remains an open question. The traditional interpretation of QFT as a theory of quantized fields described by an explicit Lagrangian does not do justice to the ubiquity of non-perturbative dualities and to the existence of intrinsically quantum theories with no semiclassical limit.
In addition to conventional methods such as lattice QFT, the desire for a non-perturbative reformulation of QFT has inspired the development of a number of successful ``bootstrap" approaches, which emphasize the set of physical observables subject to consistency and symmetry requirements. Recent years have seen a flurry of advances in the conformal bootstrap, the S-matrix bootstrap, the modular bootstrap, the cosmological bootstrap, bootstrapping string theory, and more. 

The modern view of QFT emphasizes the idea of mapping out and classifying {\it theory space}, taking fully advantage of global symmetries. These include Lorentz symmetry, conformal or superconformal symmetry, internal symmetries, and more-recently appreciated generalized symmetries. Flows through theory space which connect QFTs to themselves (or to other short-distance completions) 
can be studied using a variety of methods, including the renormalization group, truncation methods, and deformations that preserve integrability or supersymmetry. In particular, there has been significant progress in the classification of superconformal field theories (SCFTs) in various dimensions, leveraging their rigid algebraic and geometric structures and the possibility to realize many of them in string theory or by dimensional reduction of six-dimensional SCFTs. More formal approaches to QFT have also revealed a number of deep connections to mathematics, including the study of vertex operator algebras, (mock) modular forms and the moonshine program, the geometric Langlands program, and more.
 
QFTs also make deep connections to quantum gravity and string theory through holography. This connection has led to an exciting research program of constraining or reconstructing bulk physics from the boundary QFT, yielding many new insights into quantum gravity, the physics of black holes, and how they process quantum information. More generally, information-theoretic ideas such as entanglement have played a crucial role in a number of recent developments, e.g.~leading to proofs of null energy conditions, c-theorems, bounds on quantum chaos, and bolstering our understanding of holography through geometrical interpretations of entanglement. Overall, it is abundantly clear that quantum field theory is a surprisingly rich subject and that we have only just begun to elucidate its structure, how it relates to mathematics, and how it quantitatively describes the real world. 

\newpage

\section*{What is QFT?}

Quantum field theory (QFT) is the universal language of modern theoretical physics. It is the framework underlying our current understanding of the fundamental laws of nature, from the Standard Model of particle physics to early universe cosmology, and it is an essential tool in condensed matter physics, where it describes the long-distance behavior of disparate phases of matter. And yet, despite its  vast conceptual reach and  extraordinary quantitative successes, QFT is far from a mature subject. Each decade seems to bring not only novel applications of QFT, but also a rethinking of the proper way to conceptualize and define it. 
It seems undeniable that the subject has a profound unity, but we haven't yet been able to streamline how it should be taught, or to convey its full richness to  mathematicians. Indeed, while QFT continues to inspire deep developments in mathematics, it lacks itself a canonical rigorous foundation. These are indications that the conceptual work is not over.\footnote{At the outset of this report we should issue a disclaimer about references. We have chosen to restrict ourselves to only citing Snowmass 2021 white papers, which themselves do an exceptional job of mapping out the literature.}

\subsubsection*{Lagrangian QFT...}

A conventional interpretation of QFT is that it is a theory of quantized fields.
This is  how it arose historically, from the goal to quantize the electromagnetic field, and 
is still the viewpoint taken in most introductory textbooks. 
A typical textbook presentation will formulate QFT in terms of path integrals, 
which describe quantum or statistical fluctuations of ``elementary'' fields in terms of an explicit measure. One encounters expressions of the familiar schematic form
\begin{equation} \nonumber
\int [{\cal D} \varphi]\,  \exp( \frac{i}{ \hbar}  \int  d^d x \, {\cal L} [\varphi] ) \, , \quad \int [{\cal D} \varphi]\,  \exp( - \beta  \int  d^d x \, {\cal L}_E [\varphi] ) \, ,
\end{equation}
in the Minkowskian (quantum) and Euclidean (statistical) cases, respectively. The Lagrangian density ${\cal L}[ \varphi]$ or   ${\cal L}_E[ \varphi]$ is
 assumed to be a local functional of the fields. If one chooses a time slicing of the spacetime manifold, an alternative but  in fact equivalent presentation uses the canonical formalism, imposing equal time commutation relations for the field operators.

There are compelling reasons, reviewed below, to
think that this viewpoint is too narrow. Following current usage, we will use ``QFT'' more  broadly, as encompassing any model that obeys the principles of quantum mechanics, locality and relativistic invariance in a fixed spacetime background.
Nevertheless, the Lagrangian formalism has many standard and still essential uses:

\begin{itemize}

\item[(i)]  {\it Weak coupling}.
If the theory is weakly coupled ($\hbar$ or $1/\beta$ small),
it can be understood at least formally as a small perturbation around free field theory.
A free field theory is described by a solvable Gaussian path integral;  equivalently,  it corresponds in the canonical formalism to an infinite collection of harmonic oscillators, interpreted as the excitation modes of freely propagating particles. The effect of interactions (non-Gaussian terms) is introduced in a systematic perturbative expansion, expressed pictorially as a sum over Feynman diagrams. More generally, the path integral formulation
is the starting point  of semiclassical saddle expansions around non-trivial field configurations, such as instantons in gauge theories.

The perturbative expansion is at best asymptotic and generally insufficient to give an unambiguous definition of a QFT. There are however certain situations (so far restricted to a few low-dimensional models) where
(the full set of) perturbative and semiclassical data
can be leveraged to achieve a non-perturbative definition, using the mathematical machinery of  ``resurgence''.

\item[(ii)] {\it Effective field theory}.
In the modern paradigm, there are no ``renormalizability'' restrictions on the  local  interactions of a Lagrangian, in fact one is instructed to consider the most general Lagrangian
that captures all  low-energy degrees of freedom and is
compatible with the desired global symmetries. Such a general Lagrangian provides an ``effective''  description 
 up to energies $E$ smaller than a cut-off scale $M$, where new physics or strong coupling sets in. Crucially, this framework retains predictability, because in order to calculate observables to any given order in $E/M$, it is sufficient to keep only a finite number of local operators. Indeed,
renormalization theory organizes interactions  by power counting, assigning a scaling dimension 
to each local operator in the effective Lagrangian; 
operators with higher scaling dimensions are (technically and practically) more and more  ``irrelevant'' at low energies.

We have just sketched the philosophy of continuum effective field theory (EFT), the lingua franca of theoretical physics~\cite{Baumgart:2022yty}. Even dynamical gravity is easily incorporated into this framework; indeed we now think of the Einstein-Hilbert action (with a cosmological constant) as the leading term in an infinite derivative expansion~\cite{Goldberger:2022ebt}.
One can in fact argue (see e.g.~Weinberg's QFT textbook) that assuming the fundamental principles of quantum mechanics, relativistic invariance, and cluster decomposition, any scattering process 
of IR free particles will be described by {\it some} EFT. This demystifies the role of ``quantized fields''. From this viewpoint,   EFT is just a powerful book-keeping device to generate low-energy S-matrices with the requisite properties. 
It is however essential to distinguish between EFTs that arise as the low energy approximation to fully consistent, ``UV-complete''  QFTs, and those that do not. The principles of casuality and unitarity impose powerful consistency constraints on the space of  EFTs that {\it do }admit a UV completion.

\item[(iii)] {\it Non-perturbative path integrals.}
  Some indispensable tools in non-perturbative QFT relate closely to the standard Lagrangian approach.
   By formulating the theory on a  discretized spacetime, lattice QFT 
can provide a fully non-perturbative definition  of  the path integral. The existence of the continuum and infinite-volume limits has only been  established rigorously  for a limited class of low-dimensional models, 
but at a physics level of rigor renormalization theory provides clear criteria for this, which apply to
a much larger class. The most important examples are asymptotically free non-chiral gauge theories such as QCD. 
We refer to~\cite{Boyle:2022ncb, Davoudi:2022bnl} for a review of the vast subfield 
of practical non-perturbative  calculations in QCD.\footnote{Lattice methods can also be used to learn about the conformal windows of more general gauge theories and supersymmetric QFTs like $\mathcal{N} = 4$ Yang-Mills theory~\cite{Catterall:2022qzs}.}

Another non-perturbative use of path-integrals is when they can be evaluated exactly. This is the case  in topological field theories, which intrinsically have a finite number of degrees of freedom, and also for special observables in many supersymmetric field theories. In some situations, the technique of supersymmetric localization reduces
the infinite-dimensional functional integral to a tractable
semiclassical calculation that can often be performed in closed form in terms of a finite-dimensional integral.

\end{itemize}

\subsection*{...and its discontents}

There are many indications that the formulation of QFT in terms of ``fundamental fields" and Lagrangians is inadequate or incomplete: 

\begin{itemize}
     \item[(i)]
        Even at weak coupling,
     there often exist vastly more  efficient calculational methods than the standard perturbative expansion in Feynman diagrams~\cite{Bern:2022jnl}. Examples are on on-shell recursion relations and unitarity methods. What's more, perturbative amplitudes display surprising geometric structures not manifest in the Lagrangian. 
 \item[(ii)]
        The situation only gets worse at 
     strong coupling, where Lagrangians are generally not useful. (A  notable exception
     is numerical lattice simulations, which however come at the expense of breaking many symmetries and requiring significant computational resources.)  Exactly solvable models provide another hint.
     Their solution is usually obtained by indirect methods (such as integrability) which do not refer at all to the microscopic Lagrangian.
    
        \item[(iii)]  A more conceptual
        reason that the Lagrangian formulation cannot be regarded as truly fundamental is the ubiquity of non-perturbative dualities. There is a plethora of examples where the ``same'' QFT (identified abstractly in terms of its set of observables)
        can be given different Lagrangian presentations, in terms of seemingly completely different ``microscopic degrees of freedom''. A sharp version of this phenomenon is for families of supersymmetric conformal field theories 
        parametrized by manifolds of exactly marginal couplings. There are many examples of conformal manifolds that admit multiple Lagrangian formulations, each of them valid in the neighborhood of  a distinct isolated point. The quantum theory is one, but it admits multiple ``dual'' semiclassical descriptions: when one such description is weakly coupled, the others are strongly coupled. None of them has a claim to be more fundamental than the others. A perhaps useful analogy is with the definition of a manifold in terms of an atlas of overlapping coordinate charts.  Even more common is the phenomenon of IR duality, where renormalization group (RG) flows emanating from different UV theories lead to the same IR physics. 
        A hallmark of duality is that what is interpreted as fundamental in one description is emergent in the other: ``elementary'' field quanta are mapped to vortices, electrically charged particles to magnetic monopoles, etc.
        We may have become used to these phenomena, but at a deep level they remain as mysterious as ever.
   
 \item[(iv)] Finally, there is a growing list of QFTs
 that are believed to not admit any Lagrangian description whatsoever. We may call a QFT ``non-Lagrangian'' if it is {not}  connected  to a Gaussian  theory
by marginal or relevant deformations, i.e.~it is not on the conformal manifold of a free theory nor can it be reached as 
 the IR endpoint of an RG flow emanating in the UV from a free (or asymptotically free) theory.  
 A non-Lagrangian QFT is thus an intrinsically quantum theory, which does not possess any semiclassical limit.
 As it is difficult to establish such a negative, some QFTs that were once believed to be non-Lagrangian have now Lagrangian realizations. 
But there are some compelling candidates for genuinely non-Lagrangian QFTs. The six-dimensional (2, 0) superconformal field theories may be the best examples.

\end{itemize}
    
\subsection*{Axiomatic frameworks}

A more abstract framework for QFT seems sorely needed, moving away from explicit Lagrangians. Axiomatic approaches to QFT in mathematical physics have a long history and remain the subject of ongoing development~\cite{Dedushenko:2022zwd, Bah:2022xfv}. There are a bewildering number of them, falling very schematically into three categories: 
\begin{enumerate}
\item[(i)]
Correlator-based axioms, such as the 
Wightman axioms for Lorentzian QFT and the closely-related Osterwalder-Schrader axioms 
for Euclidean QFT. These axiom systems emphasize correlation functions of local operators acting on a Hilbert space. They comprise a set of consistency conditions for the ``answer" (the set of observables), while being agnostic about a more ``microscopic'' definition of QFT. As such, they foreshadow the bootstrap philosophy, as we will review below.
Starting in the 1960s, several explicit examples of  simple interacting QFTs 
 in $d<4$ have also been rigorously ``constructed'', i.e.~shown to exist 
 within these axiomatic systems in terms of (suitably regulated)  Lagrangian models.
 The Clay Millennium problem of proving that four-dimensional Yang-Mills theory exists and has a mass gap in $\mathbb{R}^4$ remains wide open. 
 Scattering theory (the LSZ formalism, etc.) was also developed within this framework, leading to some interesting (but still very partial) rigorous results
 about the analytic properties of the S-matrix in theories with a mass gap.
\item[(ii)]
Algebraic QFT axioms (e.g.~Haag-Kastler or locally covariant QFT),
which emphasize the  von Neumann algebras of observables associated to local regions of spacetime. 
A peculiarity of this approach is that Hilbert spaces of states are a derived concept.
The algebraic viewpoint has proved very useful in the study of the information-theoretic properties of QFT, such as entanglement entropy.
However, its elegance and generality comes at the price 
of being very removed from more standard physics techniques.

\item[(iii)]
Functorial QFT type axioms,
which roughly view a QFT as a functor from the category of geometric bordisms to the category of topological vector space. They have been particularly developed for topological QFT  (Atiyah-Segal axioms), where this mathematically rigorous approach has been very fruitful, leading for example to the classification of fully extended TQFTs. 
Even for non-topological theories, this is an influential viewpoint. Functoriality
encodes locality, abstracting the  natural ``gluing'' properties of the path integral formulation: the spacetime manifold can be cut into pieces that talk to each other only through their boundaries, where 
Hilbert spaces of states live.

\end{enumerate}

Each of these axiom systems formalizes some salient features of QFT, but none of them appears to be fully satisfactory. 
Their mutual relations are also poorly  understood; in particular there seems to be a profound disconnect betweeen the correlator-based approaches and the algebraic viewpoint. Bridging this divide is an important open problem.

\subsection*{The idea of theory space}

A modern idea is to lift the gaze from concrete models to the {\it space} of all possible consistent theories and their interconnections.  
Wilson thought of theory
space  in terms of a cut-off statistical model (e.g.~on the lattice) with a given field content but where all possible local interactions are allowed with arbitrary couplings.
The renormalization group (RG) 
is a flow in this infinite-dimensional parameter space, specifying how couplings evolve under a change of scale towards larger distances.  Typically, the RG flow 
converges to a finite dimensional submanifold in the space of possible Lagrangians, i.e.~the scaling transformation has only a finite number of positive eigenvalues,
  the ``relevant'' deformations, while the infinitely many orthogonal directions 
  are  ``irrelevant'' at large distance. 
  Fixed points of the RG correspond to scale-invariant theories, self-similar under a change of scale, and capture the behavior of the statistical model near a continuous phase transition.
  This picture gives a compelling intuitive explanation for critical universality,
 the empirical fact that the behavior of 
 statistical models near a continuous phase transition falls into a sparse number of classes characterized by distinct critical exponents. If one wishes to discuss continuum QFTs, one should send the short-distance cut-off to zero while keeping the low-energy physics fixed.\footnote{This is not generally possible, as one may encounter a singularity in the limit. Indeed there are rigorous results that specific lattice discretizations of scalar field theory with a $\lambda \phi^4$ interaction do not admit a continuum limit in $d \geq 4$, unless the quartic coupling is renormalized to zero. 
 (Whether there might be some other non-perturbative completion of~$\lambda \phi^4$ theory in $d=4$ remains an open question.)
 By contrast, the limit is expected to exist for  asympotically free gauge theories in $d=4$ and for many more models in $d <4$.}

In modern hep-th parlance, the phrase ``space of QFTs''  refers
 to such  ``UV-complete'' theories, which comprise a small subspace of the full set of cut-off theories with generic interactions.
 A broader notion of QFT is in fact usually implied, one that is not necessarily tied to a  concrete lattice realization or Lagrangian model, but is conceived more abstractly in terms of its set of observables. As in the Wilsonian picture, this more abstract theory space is organized by the RG flow. 
 Conformal field theories (CFTs), the fixed points\footnote{It is a not-fully-understood fact that in (unitary, Poincar\'e invariant) theories, scale invariance is generically enhanced to conformal invariance, barring a few easy-to-dispense exceptions such as free Maxwell theory in $d\neq 4$. This can be rigorously shown in $d=2$ for theories with a discrete spectrum. There are a variety of still incomplete arguments in  $d=4$.} of the RG, serve as signposts in theory space. To any conformal theory~${\cal T}$
 one can assign a positive real number $f({\cal T})$, such that $f({\cal T}_{\rm UV}) > f({\cal T}_{\rm IR})$ if ${\cal T}_{\rm UV}$ and ${\cal T}_{\rm IR}$ are connected by  RG flow.\footnote{More precisely, this has been rigorously established only in $d \leq 4$. In $d=2$, the RG monotonic function is Zamolodchikov's $c$-function, which is defined all along the RG flow and is stationary (only) at  fixed points, where it coincides with the $c$ central charge. In $d=3$, it is the ``$F$-function'', which at fixed points coincides with the three-sphere partition function; while defined along the flow, it is not necessarily stationary at fixed points. 
 In $d=4$, it is the ``$a$'' Weyl anomaly  (the coefficient of the Euler density in the trace of the stress tensor in a curved background), which is in general only defined at fixed points.} This is a powerful organizing principle of theory space.

 Any CFT has a finite number of relevant deformations,  each the seed of an RG trajectory. 
  In a given concrete model, the classic QFT question is: what is the IR behavior along a chosen RG trajectory?  There is a hierarchy in the possible richness of the long-distance physics of a continuum QFT.
  If the IR theory is gapless, it will be described by another (possibly free) CFT. If it is gapped, 
  it may be  a nontrivial TQFT with topological order and long-range entanglement; it may have no topological order 
   but be endowed with nontrivial edge degrees of freedom on a space with boundaries; finally it may be a completely trivial gapped phase.

\section*{Charting theory space}

\subsection*{Bootstrap approaches}

Recent years have seen great progress in ``bootstrap” approaches to the study of theory space, where one thinks about a QFT in terms of its set of observables subject to various consistency and symmetry requirements. This is an old idea, developed for S-matrices in the 1960s and for conformal field theories (CFTs) in the 1970s.\footnote{The essential idea of retreating to observables is already in Heisenberg's original paper on quantum mechanics, which opens with the following manifesto: {\it The present paper seeks to establish a basis for theoretical quantum mechanics founded exclusively upon relations between quantities which are in principle observable.} --W. Heisenberg (1925)} The bootstrap idea led to some impressive successes in the 1980s for 2d CFTs, e.g.~the exact solution of the 2d minimal models, but did not take off for higher-dimensional theories until more recent years.

CFTs can be defined abstractly in terms of a fairly 
well-understood set of axioms, philosophically in the spirit of the Wightman and Osterwalder-Schrader axioms.
The conformal bootstrap 
formulates consistency conditions by demanding that correlation functions can be consistently expanded in all channels using the operator product expansion (OPE). 
The resulting bootstrap equations can be studied both numerically~\cite{Poland:2022qrs} and analytically~\cite{Hartman:2022zik}. 

On the numerical side, reformulating the bootstrap equations as convex optimization problems has led to numerous bounds on the observable data of unitary CFTs, along with world-record computations of scaling dimensions (critical exponents) in theories of great physical interest, including the 3d Ising and $O(N)$ vector models. Importantly, the numerical bootstrap method gives rigorous error bars and provides a systematic approach to carving out allowed ``islands" in the space of CFTs. While most of this work involves four-point functions of scalar operators, recent work has also applied the numerical bootstrap method successfully to correlation functions of fermions, conserved currents, and the stress-energy tensor. It is also a promising tool for learning non-perturbative information about gauge theories, and in particular has recently yielded precise non-planar information in 3d $\mathcal{N}=8$ ABJM and 4d $\mathcal{N}=4$ Yang-Mills theories. While recent work has produced some highly nontrivial bounds on gauge theories with minimal or no supersymmetry, obtaining robust numerical solutions of these theories remains an important goal for the numerical bootstrap.

The bootstrap equations can also be studied analytically, and in recent years a flurry of Lorentzian methods have led to analytical solutions of CFTs at large spin, inversion formulas, conformal dispersion relations, bounds on event shapes and energy distributions, and bounds on gravitational S-matrices in AdS. To highlight a few results, analyzing the bootstrap equations in a ``lightcone" limit establishes the existence of multi-twist trajectories of operators, whose dimensions can be computed in a large spin expansion. More recently these arguments have been upgraded using the ``Lorentzian Inversion Formula", which has shown non-perturbatively that these trajectories of operators can be related to an integrated double commutator and analytically continued in spin away from integer values. Such operators can be reinterpreted as non-local ``light-ray" operators, which connect closely with null energy conditions and event shape observables. 

CFT spectra can also be studied analytically at large charge via the construction of universal effective field theories valid in this regime, while correlators involving operators of large dimension can be connected in interesting ways to the eigenstate thermalization hypothesis. In 2d CFTs there has also been great progress in applying bootstrap methods to impose modular invariance on partition functions, both numerically and analytically, producing e.g.~highly nontrivial bounds on gaps in the operator spectrum. In all of these contexts there are ongoing efforts to merge the numerical and analytical bootstrap approaches together, e.g.~the construction of analytic functionals (closely related to the solution to the sphere packing problem in mathematics) and the development of numerical-analytical hybrid bootstrap algorithms. 

A related development is the recent renaissance of the S-matrix bootstrap~\cite{Kruczenski:2022lot}. The axiomatic framework is here more tentative as the properties
of the S-matrix are not completely understood in a fully non-perturbative setting, and it is necessary to make plausible physical assumptions. By revitalizing the old idea to impose analyticity, Lorentz invariance, crossing symmetry, and unitarity on S-matrices, the space of consistent S-matrices can be mapped out using both primal and dual optimization methods. These methods have been demonstrated to work well in massive 2d QFTs where they can be compared with exact integrable theories, and there are ongoing efforts to apply these methods to QFTs in higher dimensions where they have recently produced impressive constraints on both 4d pion scattering and 10d graviton scattering. Such non-perturbative approaches to S-matrices relate closely with newly appreciated positivity bounds on effective field theory coefficients~\cite{Hartman:2022zik, deRham:2022hpx} as well as with the impressive progress that has been made in perturbative computations of S-matrices. Bootstrap ideas are also now being applied directly to lattice models and matrix quantum mechanics~\cite{Blake:2022uyo}, as well as making their way to constraining correlation functions in de Sitter space, with relevance to inflation and cosmology~\cite{Baumann:2022jpr}.

As discussed above, in many theories there is a renormalization group flow which connects a UV CFT to a gapped phase in the IR. This naturally raises the question of whether the physical properties of the IR gapped phase (e.g.~the masses and S-matrices of its particles) can be reconstructed from the data of the UV CFT and the direction of the deformation. In principle this is simply a problem of diagonalizing the Hamiltonian of the perturbed theory, and the numerical treatment of this approach goes under the name of Hamiltonian Truncation (HT)~\cite{Fitzpatrick:2022dwq}. Older versions of this idea went under the names of Truncated Conformal Space Approach (TCSA) or Discrete Light Cone Quantization (DLCQ), while a more modern version called Lightcone Conformal Truncation (LCT) emphasizes the conformal basis of the UV CFT. These methods have been successfully applied to a variety of 2d QFTs (including e.g.~2d QCD with adjoint matter), and are being actively developed in 3d and 4d. Recent work has e.g.~yielded some successful computations in 3d $\phi^4$ theory. A dream of this research program is to have a new nonperturbative scheme for computing the spectrum of 4d QCD, by flowing from a UV CFT (potentially solvable from the bootstrap) such as 4d $\mathcal{N}=4$ SYM theory.

\subsection*{Generalized symmetries}

In all of these non-perturbative methods for studying theory space, symmetry is of paramount importance, as it dictates both the set of observable data as well as the constraints that are imposed on it.
One of the most defining features of a QFT is its set of global symmetries.
A global symmetry acts non-trivially on the Hilbert space of physical states and is intrinsic to the abstract QFT.\footnote{This in contrast with a so-called gauge ``symmetry'' which is just a redundancy in a particular semiclassical description of the system, generally not preserved under a change of duality frame.}
Global symmetries include spacetime symmetries (such as Lorentz symmetry, supersymmetry, conformal symmetry) as well as internal symmetries. 
There has been a growing appreciation in recent years that the standard notion of symmetries
can be generalized in various powerful ways~\cite{Cordova:2022ruw}. These include generalizations to higher-form symmetries, higher-group symmetries, non-invertible symmetries, and subsystem symmetries. (In this nomenclature, standard symmetries are referred to as zero-form invertible symmetries).

Such generalized symmetries are ubiquitous in QFTs, e.g.~the field strength tensor of U(1) Maxwell theory describes the current of a conserved one-form symmetry, while the dual field strength tensor describes the current of a conserved $(d-3)$-form symmetry. Higher-form symmetries can be  discrete or continuous, can act on each other in nontrivial ways (creating higher-group symmetries), and can potentially have 't Hooft anomalies.  Matching their anomalies along RG flows can place strong constraints on the infrared behavior of many different physical systems. They can also spontaneously break, providing clean ways to characterize different phases of matter (extending the Landau paradigm) even when there is no breaking of zero-form symmetries. For example,
a spontaneously broken one-form symmetry leads to deconfined line defects (obeying a perimeter law at large distances), while an unbroken one-form symmetry implies confinement (an area law).

Just as zero-form symmetries are naturally supported on codimension-one manifolds in spacetime, $p$-form symmetries are naturally supported on codimension-$(p+1)$ manifolds. Every group element is consequently associated to some particular (extended) topological defect in the theory. However, in general there can also be topological defects not associated to group elements of any symmetry, but which satisfy fusion algebras that are typically non-invertible. In general the fusion ``coefficients" in these algebras are not numbers but TQFTs. Such non-invertible symmetries can be identified in many theories (including the Standard Model), and are the subject of ongoing investigation. 

Another exciting development is the identification of exotic kinds of symmetries called subsystem symmetries. Such symmetries can be found in certain lattice models called ``fracton" models~\cite{Cordova:2022ruw, Brauner:2022rvf}, which have many strange properties such as an infinite ground-state degeneracy, excitations which have restricted mobility, and a surprisingly tight connection between their UV and IR behavior. Continuum field theory descriptions of some of these models have been constructed using tensor gauge fields, but other models (e.g.~the Haah code) do not yet have a compelling continuum description. Deepening our understanding of these models gives a promising path to developing physically-relevant generalizations of standard continuum quantum field theory.

The subject of generalized symmetries has been developed in parallel in the theoretical high-energy and condensed matter communities, with insights flowing in both directions. 
 A key example of this emerging dictionary is the relation between the idea of anomaly inflow in QFT and the concept of symmetry protected topological phases in condensed matter.

\subsection*{The landscape of (S)CFTs}

While in general the problem of classifying QFTs and mapping out theory space is very ambitious, significant progress has been made in charting the landscape of SCFTs~\cite{Argyres:2022mnu}, particularly when there is extended supersymmetry. 
The additional symmetry leads to rich algebraic and geometric structures, such as moduli spaces of vacua, conformal manifolds, and protected operator subalgebras.

Moduli spaces of supersymmetric vacua  carry geometric structures which depend on the spacetime dimension, the number of supercharges, and in some cases the specific vacuum branch.\footnote{To mention two canonical examples: in $d=4$, ${\cal N}=2$ theories, Higgs branches are hyperk\"ahler, while Coulomb branches are special K\"ahler.}
Moving out in the moduli space allows to use semiclassical reasoning even in strongly-coupled theories, greatly facilitating their analysis.  SCFTs in  $d=2, 3$ and $4$  can occur in continuous families parametrized by exactly marginal couplings (conformal manifolds).  As we have already mentioned, different corners of a conformal manifold often admit different semiclassical descriptions; non-perturbative dualities are thus a ubiquitous phenomenon in these classes of SCFTs. Incidentally, 
in $d > 2$, all known examples of CFTs that admit conformal manifolds are supersymmetric. Similarly, moduli spaces of inequivalent vacua (not related by a spontaneously broken continuous symmetry) have only been observed in supersymmetric QFTs. Whether these are lamppost effects or deep facts about QFT are fundamental open questions for the bootstrap program. Finally, certain classes of SCFTs admit infinite-dimensional subsectors of protected operators, endowed with rigid algebraic structures. There is an intricate interplay between these algebraic structures and the geometry of the moduli space of vacua.

Superconformal algebras were classified by Nahm and 
exist for $d \leq 6$. 
Nevertheless, general power counting arguments based on Lagrangians and the Wilsonian RG suggest that
no interacting QFT should exist above four dimensions.
The discovery in the 1990s  of large classes of
interacting SCFTs in $d=5$ and $d=6$ (from various limits of string theory where dynamical gravity is decoupled) came thus as a real surprise,
one which has significantly enlarged the boundaries of QFT. 

It remains the case that interacting models tend to be sparser in higher dimension. 
Of special significance are the maximally supersymmetric theories\footnote{While there exists $d=6$, $(n, 0)$ superconformal algebra for any $n$, sensible QFT models are only possible for $n \leq 2$. Similarly, in $d =4$ the number of supercharges is restricted to ${\cal N} \leq 4$ and in $d=3$ to ${\cal N}\leq 8$. There is a unique superconformal algebra in $d=5$.} in the highest dimension, namely the six-dimensional  $\mathcal{N} = (2,0)$ theories, which are conjectured to admit a simple ADE classification.  
A possible classification scheme of 6d $\mathcal{N} = (1,0)$ SCFTs can be obtained via F-theory. 
Starting with the 6d theories, one can obtain large classes of SCFTs in dimension $6-n$ by compactification on $n$-dimensional manifolds $X_n$. This often leads to 
a neat geometric picture for non-perturbative properties of the lower-dimensional SCFTs.

A large class of 5d SCFTs can be obtained by compactification of the 6d SCFTs on a circle, or more generally via M-theory on non-compact singular Calabi-Yau threefolds.  An important open question (in general dimension) is whether {\it all} $\,$SCFTs can be realized in various limits of string theory. 

Dimensional reduction of the 6d (2, 0) SCFTs on  a Riemann surface yield the 4d $\mathcal{N}=2$ SCFTs ``of class S''. The complex structure moduli space of the Riemann surface is identified with the conformal manifold of the 4d SCFT. Different pairs-of-pants degenerations
 of the Riemann surface correspond to different semiclassical limits of the SCFT, where a gauge group becomes weakly coupled.
A special case is compactification on the torus, which yields ${\cal N} =4$ super Yang-Mills  theory; the complexified gauge coupling $\tau$ is identified with 
the complex modulus of the torus, and S-duality  reinterpreted geometrically as modular invariance. 
More generally, in
all known examples of 4d ${\cal N}~=~2$ conformal manifolds, the exactly marginal parameters
arise as holomorphic gauge couplings, i.e.~from gauging a subgroup of the global symmetry
group of a collection of isolated SCFTs.
All Lagrangian examples and all theories of class ${\cal S}$  are of this type,
with the isolated SCFTs being just a set of free hypermultiplets in the Lagrangian case, and
more general “matter” SCFTs in the class~${\cal S}$ case.

While the list of 4d ${\cal N}=4$ SCFTs is believed to be exhausted by ${\cal N}=4$ super Yang-Mills theories (though a proof is lacking), 
charting the space of general 4d ${\cal N}=2$ SCFTs remains a difficult open problem. A fruitful approach is to focus on the geometry of the Coulomb branch. The complexity of the task increases with the dimension of the Coulomb branch, known as the ``rank'' of the theory. A full classification has been 
achieved for rank one, and there are partial results for rank two. In a seemingly orthogonal direction, 
some partial progress can be made by studying the vertex operator algebras which are known to arise in the cohomological reduction of 4d $\mathcal{N} \geq 2$ SCFTs; these algebraic structures are more closely related to the geometry of the Higgs branch, though there are hints that the Coulomb-based and VOA-based approaches are in fact deeply connected.
There also exist 4d SCFTs with ${\cal N}=3$ supersymmetry. 
Examples have been obtained in string theory using the S-fold construction. Charting the space of ${\cal N}=3$ SCFTs will  be an ideal playground for different  classification approaches.

After further dimensional reduction one can obtain a large class of 3d SCFTs with $\mathcal{N} \geq 4$, many of which can be related by mirror symmetry. There is a general proposal for the classification of 3d $\mathcal{N} \geq 6$ SCFTs based on reflection groups, but charting the space of 3d $\mathcal{N} = 4$ SCFTs is a wide open problem. There is a sprawling list of  constructions of SCFTs with even less supersymmetry, known to live in a rich web of dualities. While their classification remains out of reach, many of these are interesting targets for the bootstrap approaches discussed above.

Our understanding of the landscape of non-supersymmetric CFTs is even more primitive. In $d=3$ there is a zoo of Lagrangian CFTs that can be reached via RG flows from theories of interacting scalars and fermions. There are also many known constructions of CFTs via (Abelian and non-Abelian) gauge and Chern-Simons theories coupled to charged matter. In $d=4$ there are many similar constructions of gauge theories coupled to matter, e.g.~the conformal window of QCD and its many variations. Establishing the precise extent of the conformal windows of gauge theories in $d=3,4$ is still an important open problem in QFT, though we have learned much from lattice methods (and are beginning to get hints from bootstrap studies). Perhaps even more exciting is the question of whether there is a whole additional landscape of CFTs in $d=3$, $4$ which are not smoothly connected to Lagrangian descriptions. 

The landscape of 2d CFTs may superficially
appear under better control; indeed the $\mathfrak{so}(d+1,1)$ conformal algebra enhances
in $d=2$ to the infinite-dimensional Virasoro algebra. 
In some theories there can be even larger algebraic structures, e.g.~$\mathcal{W}$-algebras or super-Virasoro algebras. Rational 2d CFTs contain a finite number of representations of some chiral algebra and are often exactly solvable, leading to classifications such as the ADE classification of the Virasoro minimal models with central charge $c<1$, as well as supersymmetric extensions. On the other hand, the landscape of non-rational 2d CFTs is almost completely uncharted. For example,
while RG intuition suggests that non-rational unitary CFTs with no extended chiral algebra and a discrete spectrum should be plentiful,
no single  concrete example has been constructed. As discussed above, general 2d CFTs can be constrained using conformal or modular bootstrap methods, and we can obtain additional information about 2d SCFTs via studies of their BPS spectra and moduli spaces. Charting out the full extent of the CFT theory space, both for CFTs in $d=2$ and $d>2$, will remain an important goal in the coming decades.

\subsection*{Pushing the boundaries of QFT}

It is also important to understand the boundaries of what it means to be a QFT and how they can be pushed in physically-interesting directions. 

A fairly mild extension of the framework discussed so far is to consider QFTs in curved spacetimes or on nontrivial geometries -- a subject which has a long history and makes close connections to the physics of quantum gravity, black holes, and holography~\cite{Bousso:2022ntt}. Putting a QFT in a nontrivial geometry can allow one to probe aspects of the theory which are not captured by correlation functions of local operators in flat space. E.g.~the physics of a QFT on $\mathbb{R}^{d-1} \times \mathbb{S}^1$ relates to its highly nontrivial thermal properties, while more general compactifications lead to intriguing relations between QFTs in different dimensions. The study of QFTs in Anti de Sitter (AdS) space relates directly to conventional holography (see below), while the study of effective field theories (and their UV completions) in de Sitter space plays an essential role in modern theoretical cosmology~\cite{Flauger:2022hie, Cabass:2022avo} and the cosmological bootstrap program~\cite{Baumann:2022jpr}. Putting QFTs in curved spacetimes often brings with it a number of new subtleties including infrared divergences, the choice of vacuum state and initial conditions, the breakdown of semiclassical expansions, and so on.

QFTs can also be studied in the presence of defects or boundaries. 
The allowed spectrum of extended operators or defects\footnote{Whether one thinks of an extended object as an operator or a defect depends on whether it is localized or extended in the time direction. In Euclidean QFT, the two notions are essentially indistinguishable.} provides a refinement in
the classification of QFTs: two models may be identical as far as correlators of local operators are concerned, but differ in their sectors of extended operators. For example, in a gauge theory the set of allowed Wilson and 't Hooft line operators is sensitive to the global structure of the gauge group, while local operator data  depend only on the Lie algebra.
Degrees of freedom described by lower-dimensional QFTs can live on defects or boundaries, a fact which plays an important role in both condensed matter and high energy contexts. The study of supersymmetry-preserving defects (and the theories which live on them) is an active  area of research. The classification and study of boundary conditions which preserve conformal symmetry also remains an important open problem which can be tackled using bootstrap-like approaches.

A vast and important subject (largely outside the scope of this report) is the study of  non-relativistic QFTs, which have of course many applications in the real world~\cite{Brauner:2022rvf}. Prominent examples are  EFTs describing Fermi surfaces and non-Fermi liquids, EFTs for hydrodynamics, continuum descriptions of lattice models which violate Lorentz invariance, and so on. There are also many critical points described by non-relativistic CFTs, which can be constrained by a non-relativistic analog of the conformal algebra called the Schr\"odinger algebra. While some important foundational work on these theories has been done, nonperturbative approaches like the bootstrap are not yet very well developed.

A more radical departure from the conventional framework  is to consider different ways to introduce non-locality into QFTs. One intriguing example is the study of non-commutative deformations of QFTs. Little string theories are another class of interesting non-gravitational theories which emerge in certain limits of string theory but retain various non-local properties, such as T-duality. Despite being defined by local lattice interactions, the fracton models described above also contain a UV/IR mixing reminiscent of non-local physics.

Another intriguing example that requires to broaden our definition of theory space is the irrelevant $T\bar{T}$ deformation of 2d CFTs~\cite{Flauger:2022hie}, which is a kind of  integrable flow that allows one to follow a theory up towards the UV, where it implies a breakdown of locality  (resembling a gravitational theory). S-matrices and holographic dual descriptions of such deformed theories have also been studied in detail,  using the language of Jackiw-Teitelbaum (JT) gravity.\footnote{JT gravity has also attracted considerable attention for describing aspects of the Sachdev-Ye-Kitaev (SYK) model, which is an intriguing low-dimensional toy model for quantum chaos and holography~\cite{Bousso:2022ntt, Blake:2022uyo}.}

Finally, an important way of pushing the boundaries of QFT comes from statistical mechanics, where there are many examples of physically interesting Euclidean QFTs which do not obey reflection positivity. These structures would be missed if one focused exclusively on unitary Lorentzian QFTs (and their Euclidean continuations). Non-unitary CFTs which occur at complex couplings (complex CFTs) can lead to RG flows with walking behavior, and a plausible interpretation of the bottom of conformal windows is that a pair of unitary CFTs merge and annihilate into a pair of complex CFTs. Further, non-unitary CFTs may play an important role in celestial holography and in describing holographic duals of QFTs in de Sitter space.

\section*{Connections}

Quantum field theory, sitting at the heart of theoretical physics,  has manifold deep connections with quantum gravity, quantum information science, and mathematics.

\subsection*{Holography}

Striking and powerful relations between quantum field theories and quantum gravity have been revealed through various versions of holography. The AdS/CFT correspondence, which connects quantum gravity or string theory in Anti de Sitter space to a boundary CFT, has a large number of formal and phenomenological applications~\cite{Agrawal:2022rqd, Harlow:2022qsq, Fox:2022tzz} which we will not strive to review here. 

However, let us highlight one important goal of this research program which connects closely to the rest of this report -- to understand how to reconstruct or constrain the properties of the bulk spacetime using the boundary QFT. An example of this is the idea of ``bootstrapping string theory"~\cite{Gopakumar:2022kof} -- by applying (numerical or analytical) bootstrap constraints on boundary correlators one can extract interesting information about the dual gravity theory, e.g.~the coefficients of higher-derivative interactions, bounds on gaps in the spectrum, whether there are AdS effective theories that do not admit UV completions, the forms of string amplitudes and worldsheet interactions, and more. 

In supersymmetric theories this program can be carried further, and precise connections can be made in e.g.~the planar limit of 4d $\mathcal{N}=4$ supersymmetric Yang-Mills theory where the QFT becomes integrable~\cite{Gopakumar:2022kof,Bah:2022xfv}. Integrability of the QFT can be used to compute various observable quantities, such as the anomalous dimensions of single-trace operators, to all orders in the 't Hooft coupling. In the AdS/CFT correspondence, this can be related to integrability of the string worldsheet. Exciting progress has been made in recent years at extending these computations in various directions: to other observables such as structure constants, away from the planar limit, to include defects and other deformations, and to other QFTs such as 3d ABJM theories.

Another intriguing version of holography which has emerged recently is celestial holography~\cite{Pasterski:2021raf}, which aims to describe the S-matrices for gravitational scattering in asymptotically flat space in terms of a theory with 2d conformal symmetry living on the celestial sphere. This ``celestial CFT" encodes the asymptotic symmetries of the gravity theory and gives a new way to think about its soft theorems, observable memory effects, and how to reformulate the S-matrix bootstrap for gravity.

\subsection*{Information theory}

Holography has also yielded many new insights into the properties of black holes and how they process quantum information~\cite{Bousso:2022ntt}. More generally, emphasizing information-theoretic properties has proven to be an extremely fruitful approach to learning about QFTs~\cite{Faulkner:2022mlp, Catterall:2022wjq}. Focusing on concepts such as entanglement and the spreading of information has yielded a plethora of insights into quantum field theory, and is closely aligned with the algebraic QFT point of view. Various path integral methods have been developed for directly computing entanglement entropy and related quantities in QFTs, while in holographic theories it can be computed geometrically using the Ryu-Takayanagi formula and its generalizations. While there are sometimes ambiguities in the definition of entanglement entropy, a more rigorous approach that has proved useful comes from Tomita-Takesaki modular theory and the idea of modular flow. 

The information-theoretic perspective has led to proofs of the irreversibility of renormalization group flow in various dimensions, including the c-theorem in 2d, the F-theorem in 3d, and the a-theorem in 4d. Interestingly, there is a dichotomy between even and odd dimensions -- while in $d=2, 4$ the quantity that decreases under RG can be related to a Weyl anomaly and extracted from local correlators,  in $d=3$ it is only related to a nonlocal  observable. This pattern is believed to persist in higher dimensions,
 but there are no proofs. These are indications that
a unified conceptual picture is still   lacking. 

Information-theoretic ideas have also yielded new important insights into the non-existence of global symmetries in quantum gravity. They have additionally played an important role in the classification of topological phases and have inspired the development of numerical tensor network methods~\cite{Meurice:2022xbk}, which emphasize entanglement and give an attractive alternative to traditional Monte Carlo approaches. 

Another fruitful direction has been the study of inequalities called null energy conditions, which can been proved using information-theoretic techniques and place highly nontrivial constraints on QFTs. A related direction is to understand the dynamics and spreading of quantum information, closely related to transport properties of QFT~\cite{Blake:2022uyo}. A particularly exciting direction has been the study of out-of-time-order correlation functions, which can be used to probe scrambling and bound the onset of quantum chaos, while at the same time connecting with more traditional QFT observables.

\subsection*{Mathematics}

Finally, many important directions of research sit right at the interface between  QFT and mathematics~\cite{Argyres:2022mnu, Bah:2022xfv}. Let us mention a few highlights.

Direct connections exist between the SCFT classification program and the theory of canonical singularities in algebraic geometry. Related to this is the rich mathematics of enumerative invariants, e.g.~Donaldson-Thomas invariants, which count certain wrapped brane states in string theory. The rich mathematics of quantum integrable systems shows up in the study of supersymmetric vacua of higher dimensional theories (the Bethe/gauge correspondence). Ideas and methods arising from the bootstrap and holography are also being applied directly to questions in spectral geometry, e.g.~bounding Laplacian eigenvalues on various manifolds.

The study of 2d rational CFTs and associated vertex operator algebras has additionally proven to be a fruitful playground for making striking connections between physics and mathematics. Intriguing generalizations of these algebraic structures can be obtained by applying topological twists to higher-dimensional theories. More generally, topological quantum field theories (TQFTs) in any dimension remain a rich arena for bridging the gap between mathematics and physics. TQFTs are the most mathematically developed QFTs and have well-understood axiomatic frameworks (though there is a certain class of TQFTs, those of cohomological type, which do not fit into existing systems). TQFTs can live on boundaries and defects in general QFTs and play an important role in the classification of invertible phases. They make rich connections to many areas of mathematics, including topology, symplectic geometry, category theory and higher algebra, geometric representation theory, and number theory. 

Deep connections between physics and number theory often show up in the modular properties of partition functions or indices which count BPS states. An exciting example of this is the moonshine program~\cite{Harrison:2022zee}, which has revealed striking relations between modular (or mock modular) forms and the representations of various finite groups. These seemingly disparate areas of mathematics can be connected through the physics of 2d conformal field theories, vertex operator algebras, and associated constructions in string compactifications. Various moonshine relations are now known to exist, e.g.~Monstrous moonshine, Mathieu moonshine, umbral moonshine, and penumbral moonshine. Yet further deep relations between number theory, algebraic geometry, and quantum field theory come from the geometric Langlands program. Identifying new physical understandings of these rich mathematical relations is bound to be a fruitful direction for many years to come.

\section*{Outlook}

Quantum field theory has come a long way since its origin in attempts to quantize the electromagnetic field. It has revealed rich mathematical structures with great physical relevance, and yet we still don't understand the best way to conceptualize it. Nevertheless, recent years have produced an abundance of powerful new ideas for how to think about QFTs, calculate with QFTs, and relate QFTs to themselves and other physical (and mathematical) systems. 

One may worry that we are still missing a deep idea about how to properly characterize what it means to be a QFT. E.g.~one may speculate that the apparent failure of conventional ideas about naturalness~\cite{Craig:2022uua} in (extensions of) the Standard Model provides a hint that we are not yet thinking about the relation between UV and IR behavior of QFTs in the correct way. We now know that there are numerous nontrivial constraints on IR effective field theories coming from UV consistency, closely analogous to the many bootstrap constraints we understand to hold on the low-dimension sector of CFTs. 
Furthering our understanding of these connections, in particular the fundamental role played by causality, will surely lead to a deeper understanding of QFT.

As we have discussed, there are a number of promising avenues for a reformulation of QFT that are being actively pursued, including axiomatic approaches, bootstrap approaches focused on observables subject to consistency conditions, approaches centered on information-theoretic quantities, and definitions of QFTs via holography or limits of string theory. Each of these approaches has distinct advantages, but also seem to lose the close connections to locality and ``microscopic'' degrees of freedom built into the Lagrangian and path integral descriptions. The ubiquity of non-perturbative dualities
could be considered another essential feature of QFT to which 
the more abstract approaches do not do full justice.

Perhaps a new formalism will someday be found which retains the essential features of all these approaches. It is clear that the task of charting out theory space, bolstered by new understandings of symmetries and modern computational tools, will remain an important and fruitful endeavor for decades to come.

\section*{Acknowledgements}
We thank  Nathaniel Craig, Csaba Csaki, and Aida El-Khadra for spearheading the efforts of the Snowmass Theory Frontier. We are grateful to all of the incredible community members who contributed to Snowmass 2021 -- by writing white papers, participating in the conferences and on-line forums, and  providing feedback to our draft. 

DP is supported by Simons Foundation grant 488651 (Simons Collaboration on the Nonperturbative Bootstrap) and DOE grant no.\ DE-SC0017660. LR is supported in part by NSF grant PHY-1915093 and by Simons Foundation grants 397411 (Simons Collaboration on the Nonperturbative Bootstrap) and 681267 (Simons Investigator Award).

\appendix 

\bibliographystyle{JHEP}
\bibliography{refs}

\end{document}